\documentstyle[12pt,aaspp4]{article}

\def\be{\begin{equation}}
\def\ee{\end{equation}} 

\def\al{\alpha}
\def\th{\theta}
\def\vth{\vartheta}
\def\lam{\lambda} 
\def\frw{Friedmann-Robertson-Walker } 
\def\Dl{\Delta} 

\def\Om{\Omega}

\begin{document} 

\title{Transition from Clumpy to Smooth Angular Diameter Distances}
\author{Eric V.~Linder} 
\affil{Department of Physics and Astronomy, University of Massachusetts, 
Amherst, MA 01003; linder@kea.phast.umass.edu}

\begin{abstract} 
Distance relations in a locally  inhomogeneous  universe are expected to 
behave like the Dyer-Roeder solution on small angular scales and the 
Friedmann-Robertson-Walker solution  on large  angular scales.    Within 
a simple compact clump model the transition between these  asymptotic 
behaviors is demonstrated and quantified.  The  redshift dependent 
transition scale is of order a few arcseconds; this implies it should 
have little influence on large angular scale cosmological tests such as 
the volume-redshift relation but possibly significant effects on arcsecond 
angular diameter measurements of radio galaxies and AGNs.  For example, 
at $z=2$ on arcsecond scales a clumpy flat universe mimics the angular 
diameter distance of a smooth $\Omega=0.27$ model. 
\end{abstract}

\section{Introduction}
Light  propagation through our universe is generally treated in two separate 
regimes: either within the smooth Friedmann-Robertson-Walker model or 
within the gravitational lensing model of density inhomogeneities (see, 
e.g., \cite{sch92}). 
In the actual universe both local inhomogeneities and global homogeneity 
exist, so each acts as an asymptotic description under the appropriate 
conditions.  Along with image amplifications, distortions, and time 
delays, the source distance-redshift 
relation is also affected by inhomogeneities.   While many researchers 
discuss the distance relations in the 
presence of inhomogeneities, e.g.~Futamase \& Sasaki (1989), Watanabe \& 
Tomita (1990), and how ignorance of clumpiness can affect determination 
of cosmological parameters by distance tests, e.g.~Linder (1988a), 
Hadrovi\'c \& Binney (1997), 
the question of the nature of the transition region between 
the two regimes is not addressed.  It is of importance to understand 
where this transition occurs so as to use the two limiting cases only 
where they are valid approximations. 

For almost all astrophysical applications of light propagation geometric 
optics holds, where 
the curvature or inhomogeneity scales 
are much greater than the wavelength of the  electromagnetic radiation, 
so the radiation can be treated as a  beam or bundle of light rays. 
The focusing of the bundle due to the spacetime geometry determines 
the angular diameter distance $r$ by 

\be 
d^2r/d\lambda^2=-({\cal R}+|\sigma|^2)\,r, 
\label{eqn:beam} 
\ee
where $\lambda$ is the affine parameter measuring the path length, 
${\cal R}=(1/2)R_{\mu\nu}k^\mu k^\nu$ 
is the Ricci contribution 
of  the gravitational focusing from matter within the beam, and 
$\sigma$ is the Weyl contribution of gravitational shear from matter 
outside the beam.  Here $R_{\mu\nu}$ is the Ricci tensor and $k^\mu$ the 
photon four momentum. 

To obtain a universal, i.e.~isotropic, distance relation, equation (\ref
{eqn:beam}) is generally 
solved under the  ``average path'' assumption (\cite{dye73}) 
for an infinitesimal light beam, e.g.~a single ray.  This posits 
that the matter density 
along the line of sight is given by the global average density times a 
uniform smoothness parameter $\al$, measuring the degree of small 
scale inhomogeneity  or clumpiness, and that the shear vanishes from 
global homogeneity.  This approximation substitutes for detailed  
knowledge of the locally   inhomogeneous metric, or physical conditions, 
along  the light path, which we generally  lack.  Given that ignorance 
some such  effective model must be adopted, with the  main caution 
being insurance that a  ``typical''  path  is indeed characteristic 
of the average. 

As the light beam subtends larger and larger solid angles, at some point 
the inhomogeneities should be smoothed over and we can legitimately 
calculate distances within the \frw cosmology.  The question arises 
how to connect these two asymptotic behaviors with a reasonable, 
preferably simple analytic model.  (One can of course use numerical  
ray shooting within a specific pattern of inhomogeneities to compute 
the distance relations but they will be relation{\it s}, not one single 
universal relation as desired.)  Of particular 
observational interest is how large is the  transition angle, as many 
cosmological tests such as the number-redshift and magnitude-redshift 
relations are sensitive to the precise distance measure. 

\section{Smoothing} 
The standard model for density inhomogeneities when calculating 
cosmological distance relations is the  Dyer-Roeder prescription:  
distribute randomly a fraction $1-\al$ of the total matter density in 
clumps, point masses $M$ with constant comoving number density $n_0$.  
Light rays passing 
too close to a clump will be appreciably  gravitationally influenced 
(spatially deflected, cross sectionally sheared, intensity amplified) 
and so will  be recognized as something special and not treated as 
being on a typical line of sight.  That leaves those paths avoiding 
clumps, but their neighborhoods only possess a matter density of  
$\al\rho$ where $\rho$ is the global average matter density.  Thus, 
an ``average'' path does not feel the full \frw (FRW) density and the 
calculated angular diameter distance $r(\al,z)$ will differ  from the FRW 
relation $r_{FRW}=r(1,z)$.  In this case  ${\cal R}=4\pi(1+z)^2\al\rho$ 
where $z$ is the source redshift.  

[One can also generalize the  model to redshift dependent 
clumpiness $\al(z)$, modeling the effects of evolving inhomogeneity, 
clump mass, or number density (\cite{lin88b}).  Note that this brings 
up an interesting point.  If we lived in a local inhomogeneity, either 
a bubble of different density or different clumpiness, then even though 
the volume averaged universe would appear Friedmann, our observed 
distance-redshift relation would not be the same as in that volume 
averaged universe.  It does not asymptotically approach the Friedmann 
result even at distances much greater than the bubble size.  
I.e.~the observer is in a special position and  
volume rather than angle (line of sight) averaging may give misleading 
results; one cannot invoke an ergodic homogeneity because our position 
as observers selects a unique location.  Such an isotropic inhomogeneity 
would lead to an apparent conflict between cosmological parameters derived 
from dynamical quantities and those involving the distance-redshift relation 
(in preparation).] 

Given the Dyer-Roeder ansatz, one can see that as a light beam subtends a 
larger solid angle,  the probability increases that it will include 
a clump.  This raises the matter density within the beam and when 
the bundle is broad  enough it will feel the full FRW  density.  To 
investigate this  transition consider a  beam with half angle $\th$ 
at the observer, 
propagating from a redshift $z$.  Let each clump own  a conical 
volume $V_c$ given by 

\be 
\int n\,dV_c=1, 
\label{eqn:volc} 
\ee
which extends from the observer out to the source or survey 
depth $z$ with half angle $\th_c$.  Here $n$ is the proper number 
density; note the subscript $c$ on the proper volume element $dV$ 
denotes ``clump'' not ``comoving''. 

The probability that the light beam will include the clump is given 
by the Poisson process 

\be
p(\th)=1-e^{-\tau(\th)}, 
\label{eqn:prob} 
\ee
where the optical depth $\tau=V(\th,z)/V(\th_c(z))$.  That is, 
on average a beam covering $V_c$ has roughly  unity probability 
for including the clump (really $1-e^{-1}$  because of statistical 
fluctuations in the number density). 

Although  as the beam expands, i.e.~$\th$ increases, the  matter within  
the beam actually  jumps in a step  function, as  the next step in our 
ansatz we simulate the global averaging  by choosing an effective 
smoothed  density $\rho_{eff}=\bar\al\rho+\rho(1-\bar\al)\,p(\th)$, where 
$\bar\al$ is the global smoothness parameter, i.e.~that felt by an 
infinitesimal ray.  So rather 
than having no clump within the beam then suddenly all of it, we smoothly  
add  its contribution according to the probability for inclusion $p(\th)$.  
This is the main approximation we make in order to talk about a universal 
distance function.  It seems physically reasonable and is well behaved 
mathematically. 

The effective smoothness parameter is now 

\be
\al(\th,z)=\bar\al+(1-\bar\al)\biggl[1-\exp\bigl\{-V(\th,z)/V[\th_c(z)]
\bigr\}\biggr]. 
\label{eqn:alv}
\ee
The volumes are given by 

\begin{eqnarray} 
V(\th,z)&=&\int_z dr_p\int_\th dA=\int_z dr_p\int_\th d\omega 
\,r^2[\al(\vth)]\nonumber\\ 
&=&2\pi\int_z dr_p\int_0^\th d\vth\,\vth r^2[\al(\vth)]\equiv
\pi\th^2g(z,\al), 
\label{eqn:volr}
\end{eqnarray} 
where $r_p$ is the  proper distance along the line of sight, $A$ is 
the transverse area on the  sky, and $\omega$ the angular area on the 
sky.  

Substituting equation (\ref{eqn:volr}) into (\ref{eqn:alv}) yields 

\begin{mathletters} 
\begin{eqnarray} 
\al(\th)&=&\bar\al+(1-\bar\al)\biggl[1-\exp\bigl\{-(\th/\th_c)^2
[g(z,\al)/g(z,\al_c)]\bigr\}\biggr], \label{eqn:aex}\\  
\al_{apx}(\th)&\equiv&\bar\al+(1-\bar\al)\bigl[1-\exp\{-(\th/\th_c)^2\}
\bigr], \label{eqn:aapx} 
\end{eqnarray}
\end{mathletters}
plotted in Figure 1, where $\al_c\equiv\al(\th_c)=1-e^{-1}(1-\bar\al)$. 
We  adopt $\Omega=1$, which would cause the greatest deviation of the 
clumpy vs.~smooth  FRW model. 
Note that the approximation in (\ref{eqn:aapx}) (solid curve) has removed 
the need for 
recursion in determining $\al(\th)$.  Because the ratio of $g$'s is 
insensitive to $\al$ the approximation 
is excellent: the fully  recursed $\al$ 
differs from  $\al_{apx}$ by at most $11\%$ for the extreme case of 
$\bar\al=0$, $z=3$ (and is more typically $<1\%$) -- and this occurs in a 
region of parameter space such that the distance $r(\al)$ is never more  
than $1\%$  from $r(\al_{apx})$. 

The resulting $\al(\th)$ has the pleasing properties of monotonicity, 
simplicity, and the proper asymptotic behaviors:  as $\th$ increases 
the parameter goes from $\al(\th\ll\th_c)=
\bar\al$ to $\al(\th\gg\th_c)=1$, i.e.~we have effectively introduced  
an averaging procedure  that provides the Dyer-Roeder clumpy  universe 
result at small angles, the FRW result at large angles, and defines 
the transition. 

To find the transition angle $\th_c$ combine equations (\ref{eqn:volc}) 
and (\ref{eqn:volr}) to get 

\be 
1=2\pi n_0\int_0^{\th_c}d\th\,\th\int_z dr_p\,(1+z)^3r^2[\al(\th)]
\equiv \pi\th_c^2 
n_0H_0^{-3}f(z,\al_c), 
\label{eqn:thcint}
\ee 
for constant comoving clump density.  So 

\be
\th_c=\bigl[\pi n_0H_0^{-3}f(z,\al_c)\bigr]^{-1/2}. 
\label{eqn:thceq} 
\ee 
Taking $n_0=(3H_0^2/8\pi)\Omega_c M^{-1}$  where $H_0=100\,
h$ km s${}^{-1}$ Mpc${}^{-1}$ is the Hubble constant and $\Omega_c=
(1-\bar\al)\Omega$ the clump matter density in units of 
the FRW critical density, 

\be 
\th_c=1.3''\,(M/10^{12}M_\odot)^{1/2} h^{1/2}(1-\bar\al)^{-1/2} 
f^{-1/2}. 
\label{eqn:thcval} 
\ee 
At low $z$ or high smoothness $1-\bar\al\ll1$ the clump angle formally  
becomes large but this is unphysical given our interpretation of $\th_c$  
as a transition angle to FRW behavior.  A more physical transition 
angle $\Theta$ is defined as that angle where 
$r(\al(\theta))$  first differs from $r_{FRW}$ by less than 5\%. 

Figure 2 plots both $\Theta$ and $\th_c$ vs.~$z$.  For low $z$ the 
distances $r$ are 
insensitive to $\al$  (it enters at third order in an 
expansion of $r$ in $z$; see \cite{lin88b}) so even for $\th=0$ all 
$r(\al)$ are close to $r_{FRW}$.  For high redshifts $\th_c$ and 
$\Theta$ formally diverge as $\bar\al\to1$ due respectively to the 
scarcity  of clumps and to the difference in high $z$ asymptotic 
behavior between smooth and even slightly 
clumpy distances.   The growth is extremely slow, however, $\th_c
(z\gg1)\approx 25''$ for $\bar\al=0.999$ and $\Theta(z=10^3)\approx 
7''$ for $\bar\al=0.99$.  Realistically, $\bar\al$ evolves due to structure 
formation such that it approaches unity sufficiently closely (e.g. $>0.998$ 
for $z\le10^3$) that $\Theta\to0$ at high redshifts.  Thus, 
in general distant source observations with fields of view 
larger than a few arcseconds can legitimately be 
treated within the FRW  model for distances. 

As a definite example, consider the volume-redshift cosmological test 
which involves the angular diameter distance (note the luminosity 
and proper motion distances are simply related to the angular 
diameter distance by factors of $1+z$).  From equation (\ref{eqn:volr}) 
the  volume-redshift relation is 

\be 
V(\th)=V_{FRW}(\th)[g(z,\al)/g(z,1)], 
\label{eqn:volz} 
\ee 
where $\al_{FRW}=1$.  For $z<1$ the  ratio  of $g$ factors is close 
to unity,  and for $z>1$ we observe at $\th\gg\th_c$ (larger than 
arcsecond fields) and so $\al(\th)\approx1$ and again $V\approx V_{FRW}$. 
Thus clumpiness effects are negligible in this case. 

The  differential volume test is more sensitive since this does not 
possess the integration in equation (\ref{eqn:volr}) that includes those 
very low redshifts 
where $r$ is nearly independent of $\al$ (as discussed above).  Then 

\be 
\Delta V(\th)/\Delta V_{FRW}(\th)= 2\th^{-2}\int_0^\th 
d\vartheta\,\vartheta r^2[\al(\vartheta)]/r^2(1). 
\label{eqn:difv} 
\ee 
The ratio is bounded between $[r(\al(\th))/r(1)]^2$ and $[r(0)/r(1)]^2$ 
and can approach $1.25$ at $z=1$ as $\th\to0$ if 
$\al(\th)$ were unreasonably pushed to zero.  The deviation from unity 
is less than $5\%$ for all redshifts and clumpiness factors, however, 
when $\th>12''$.  
In particular, this assures us that the usual  observations can be 
analyzed properly using the FRW volume element: for example the 
Loh-Spillar (1986) fields were $7'\times10'$  at $z<0.75$.  (See 
\cite{omo90} for how the clumpy luminosity distance affects the 
analysis of flux weighted counts.) 

Clumpy and transition regimes can be important, however, for cosmological 
tests involving arcsecond scales.  Possible applications include 
observations of radio galaxy lobes (\cite{kap89}, \cite{gue96}) and 
milliarcsecond observations of active galactic nuclei (\cite{gur94}) for 
use in the angular diameter distance-redshift cosmological test.  At a 
redshift of two, for example, the difference between the clumpy and smooth 
angular diameter distances for $\Omega=1$ is $33\%$, and a clumpy flat model  
gives the same distance as a smooth model with $\Omega=0.27$. 

Figure 3 shows this trade off between clumpiness in a $\Omega=1$ model and 
low density in a smooth model.  Thus a flat but clumpy universe could be 
misinterpreted through small angular scale observations  as a lower 
density one.  In general one can match a distance in a ($\al$,$\Om$)  
universe with a ($\al'>\al$,$\Om'<\Om$)  model.  The numbers on the plotted 
curves give the angular scale of the observations in arcseconds and show 
the transition in clumpiness or alternatively density miscalculation 
from the infinitesimal case (Dyer-Roeder) to the large scale average (FRW). 

\section{Shear} 
One point remains of which we must be cautious.  The ${\cal R}$ term in 
equation (\ref{eqn:beam}) was treated by using an effective smoothed 
density but we 
neglected the shear term $|\sigma|^2$.  On large scales we expect 
the shear to average to zero due to homogeneity but this is not ensured 
at smaller scales: if the light passes far from the clump then the 
shear on the bundle should be small, but not if it passes near the clump. 

We can analyze this in terms of an effective clumpiness defined 
through equation (\ref{eqn:beam}) by 

\be 
\al_{eff}=\al-(2/3)|\sigma|^2H_0^{-2}\Omega^{-1}(1+z)^{-5}, 
\label{eqn:aeff} 
\ee  
where $\Omega$ is the ratio of the total density to the critical  
density.  Then the right hand side of equation (\ref{eqn:beam}) can be 
written simply as 
$4\pi(1+z)^2\al_{eff}\rho r$ and the previous results hold with the 
substitution of $\al_{eff}$ for $\al$.  Note that shear always decreases 
$\al_{eff}$ -- brings it further from the FRW value of unity -- 
and so increases the transition angle  $\th_c$ (more properly decreases 
the ratio $\th/\th_c$ for which $\al_{eff}$ is a given value).  Also 
note that $\al_{eff}$ can be negative, which causes no mathematical 
worries [it merely makes the usual parameter  $\beta\equiv(25-24\al)^
{1/2}>5$]. 

Alternately, we can solve the full beam equation (\ref{eqn:beam}). 
Because the shear squared source term of a point mass dies off as 
$({\rm distance})^{-4}$, we  can approximate its behavior 
as localized  
in affine parameter, say between $\lam_0$ and $\lam_0+\Dl\lam$.  
Dividing the light propagation from source at $\lam_s$ to observer 
at 0 into three regimes $(0,\lam_0)$, 
$(\lam_0,\lam_0+\Dl\lam)$, $(\lam_0+\Dl\lam,\lam_s)$, we match the values 
of $r$ and its first derivative at the boundaries.  For the most 
extreme case, $\al=0$ and $\Omega=1$, the solutions to the distance 
relations are: 
$r_1=\lam$; $r_2=C\sin 
|\sigma|\lam+D\cos |\sigma|\lam$; $r_3=A\lam+B$.  Taylor expanding 
under the assumption $\Dl\lam\ll\lam_0,|\sigma|^{-1}$ yields 

\begin{eqnarray} 
r_3(\lam)&\approx& \lam-|\sigma|^2\lam_0^2\,\Dl\lam(\lam/\lam_0
-1)\nonumber\\ 
&\approx&  r_1(\lam)-(4/25)|\sigma|^2\Dl z\,y_0^{-6}(1-y_0^{-5/2})[1- 
(y_0/y)^{5/2}], 
\label{eqn:rsig}
\end{eqnarray} 
where  $y=1+z$ and $\lam=(2/5)(1-y^{-5/2})$. 

Thus the shear only has a significant effect  on the angular diameter 
distance relation if the second term is nonnegligible.  The maximum shear 
on the beam is given in order of magnitude by $|\sigma|\sim Mb^{-2}$ where  
$b$ is the impact parameter from the mass $M$.  For $b$ equal to  the 
Einstein radius of the mass, $|\sigma|\sim H_0(r_s/r_lr_{ls})$ where 
the $r$'s are respectively measured from observer to source, observer 
to lens mass, and lens to source.  The maximum value of  the ratio of 
the second to first terms in equation (\ref{eqn:rsig}) is then 
$(H_0\lam_0)^2\Dl\lam/\lam\ll1$, 
meaning that when concerned with distances we can neglect shear for beams 
passing outside the Einstein ring of the mass. 

So we argue that we can neglect the influence of shear on distances overall 
because 1) narrow beams ($\th\ll\th_E=b_E/r_l$) would typically miss 
the Einstein ring because their probability for intersection ($\th_E^2/
\th_c^2$) is small (e.g. $\th_E^2/\th_c^2\approx\Omega_c z^2\ll1$ for 
$z\ll1$); 
2) those few appreciably sheared light beams would be recognized as 
atypical and not used for distance measures; and  3) broad beams would 
have any shear effects diluted due to those portions of the beam  
that lie far from  the clump and those symmetric (isotropic) about it.  
Therefore 
we do not expect shear to alter significantly the distance-redshift 
relation for a typical light bundle, except possibly at high redshifts 
where the transition scale $\th_c$ is below an arcsecond (and so $\th_c 
<\th_E$) and hence below the region of observational interest.  

\section{Conclusion} 
Using a simple toy model of an effective density  distribution due to 
inhomogeneities one can derive a universal angular diameter distance 
relation applicable over all angular scales.  It has the desired asymptotic 
properties: agreeing for small solid angles with the clumpy universe 
Dyer-Roeder distance, recreating for large solid angle observations 
the Friedmann-Robertson-Walker relation, and interpolating smoothly 
between them.  The transition angle depends on the type of inhomogeneities 
but is estimated to be unlikely  to exceed $10''$ for any cosmological 
test and should be of order $1''$ for any quantities involving the 
entire light propagation path between the source and observer.  Still, this 
can have significant effects on such relations as the angular 
diameter-redshift relation for radio galaxies and AGNs, for example a true  
clumpy flat universe mimicking a smooth open model, and must be looked  for 
in such cosmological tests. 

\acknowledgments 
I thank Neal Katz for useful suggestions.  This work was supported by  
NASA grants NAG5-3525 and NAG5-4064. 
\vfill\eject

\vfill\eject 

\figcaption{
The effective smoothness $\al$ is plotted vs.~beam 
size $\th$ in units of the clump angle $\th_c$.  The lower curves are 
for a completely clumpy universe, $\bar\al=0$, while the upper have 
$\bar\al=0.5$.  Solid lines illustrate $\al_{apx}$ from (\ref{eqn:aapx}), 
dotted and dashed lines use the full $\al$ from (\ref{eqn:aex}), evaluated 
at survey depths of $z=1$ and $z=3$ respectively. 
\label{fig1}}

\figcaption{
The transition angle $\Theta$ (solid curves) and 
clump angle $\th_c$ (dotted curves) in arcseconds are plotted vs.~redshift.  
The pairs of curves are labeled near their intersections by 
the clumpiness $\bar\al$.  At low redshifts the distance  
relation is close to the FRW behavior for all beam sizes while at high 
$z$ it makes a transition to the clumpy behavior for angles smaller than 
$\Theta$. 
\label{fig2}} 

\figcaption{
The clumpiness $\al$ needed for a flat universe to 
have the same angular diameter distance to redshift $z$ as a smooth 
$\Om<1$ universe is plotted.  In the conventional Dyer-Roeder model, 
for example, at $z=2$ one could match $r(z)$ for $\al=0$, $\Om=1$ by a 
FRW model with $\Om=0.27$.  In the transition model of this paper, 
though, $\al$ is effectively a function of beam size.  The numbers 
superposed on the curves give this size in arcseconds (for infinitesimal 
beam clumpiness $\bar\al=0$).  Thus observations with a $2''$  beam at 
$z=2$ in a clumpy flat universe can be misinterpreted as belonging to 
a $\Om=0.54$ FRW universe. 
\label{fig3}} 
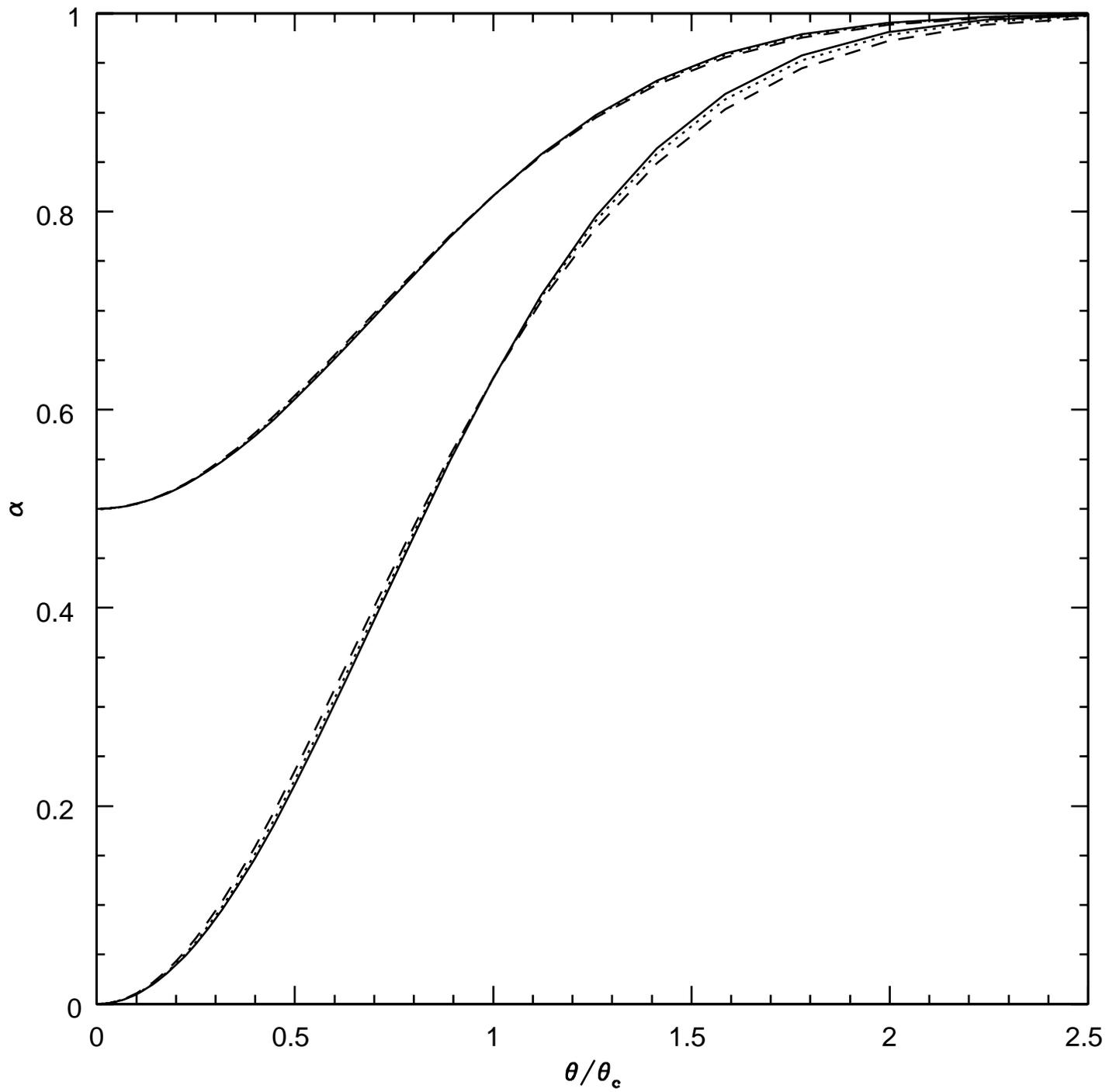
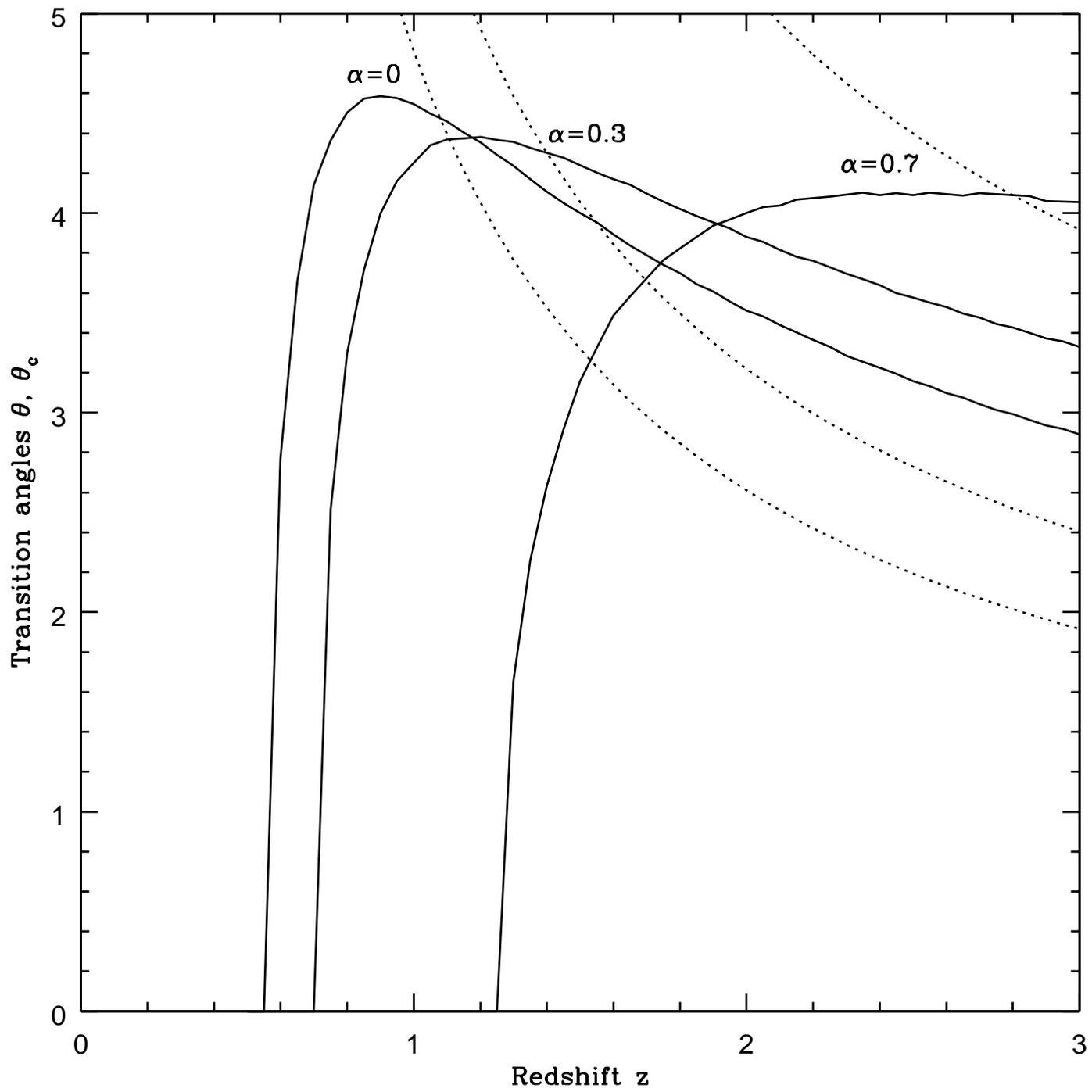
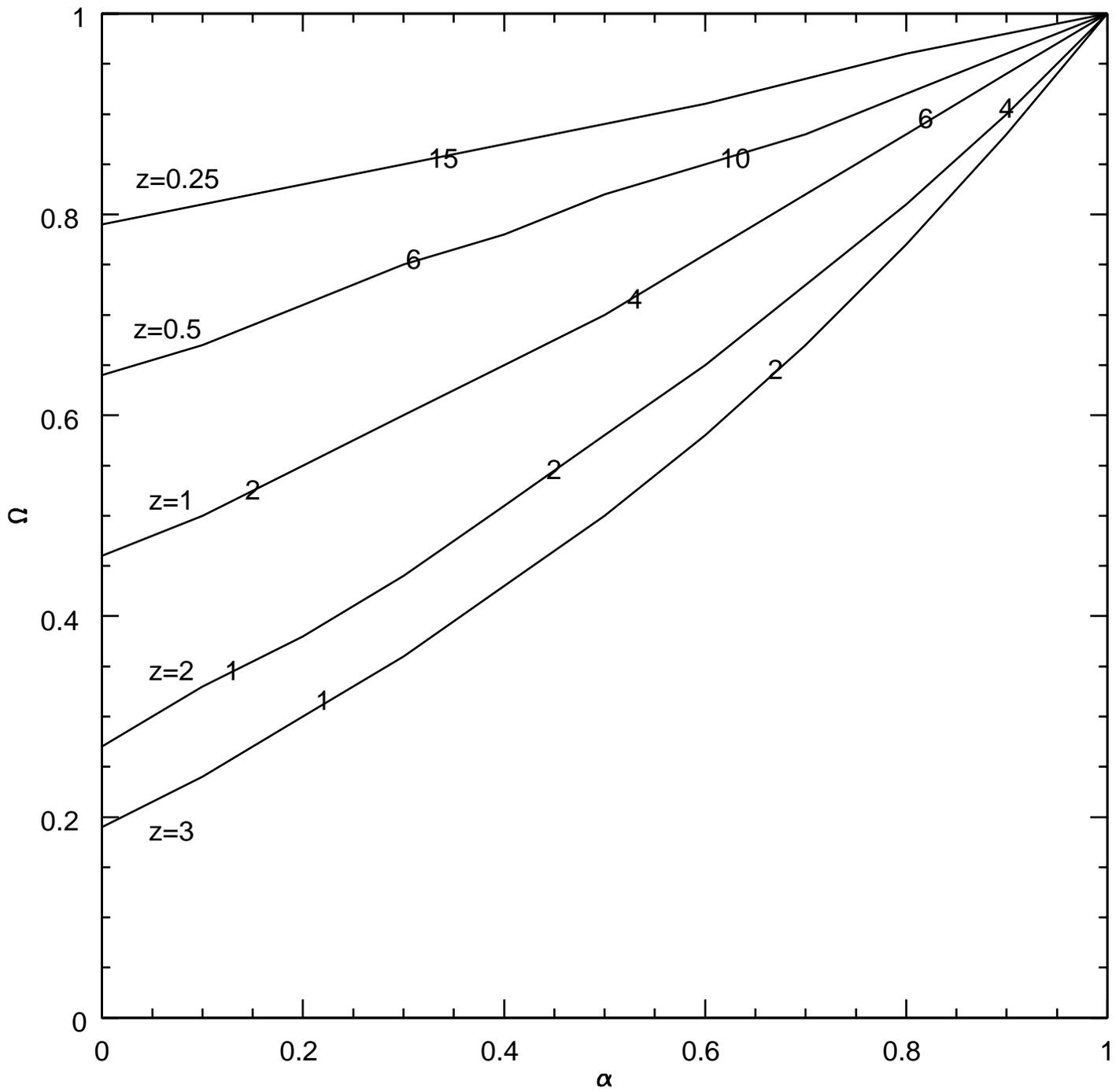
\end{document}